\newcommand{\arcsec}{{\hbox{$^{\prime\prime}$}}}
\newcommand{\kms}{km~s$^{-1}$}

\newcommand{\etal}{{et al.\/}}

\newcommand{\hei}{{He~{\sc i}}}
\newcommand{\oiii}{{O~{\sc iii}}}
\newcommand{\oiv}{{O~{\sc iv}}}
\newcommand{\ov}{{O~{\sc v}}}
\newcommand{\nev}{{Ne~{\sc v}}}
\newcommand{\nevi}{{Ne~{\sc vi}}}

\newcommand{\mgix}{{Mg~{\sc ix}}}
\newcommand{\fexvi}{{Fe~{\sc xvi}}}
\newcommand{\asa}{{ A\&A\/}}
\newcommand{\asas}{{A\&AS, \/}}
\newcommand{\apj}{{ ApJ\/}}
\newcommand{\apjl}{{ Astrophys. J. (Letters)\/}}
\newcommand{\apjs}{{ ApJS\/}}
\newcommand{\sps}{{ Sol. Phys.\/}}

\documentclass{aa}
\usepackage{graphicx}
\usepackage{color}
\begin{document}
\input epsf.sty

\title{A solar active region loop compared with a 2D~MHD model}

\titlerunning{A Solar Active Region Loop}
\authorrunning{Gontikakis et al.}
\author{\bf C.~Gontikakis,
           \inst{1}
           \,
           G.~J.~D.~Petrie,
           \inst{2}\fnmsep\thanks{Visiting Fellowship,
             High Altitude Observatory \& Scientific Computing Division,
             National Center for Atmospheric Research,
             PO Box 3000, Boulder, CO 80307-3000, USA}
            \,
             H.~C.~Dara
            \inst{1}
            \and
            K.~Tsinganos
            \inst{2}
            }  
\offprints{C. Gontikakis}

\institute{ Research Center of Astronomy and Applied Mathematics, Academy of Athens, \\
Soranou Efessiou 4,  GR-115~27, Athens, Greece
       \email{cgontik@cc.uoa.gr,edara@cc.uoa.gr}
    \and
IASA and Section of Astrophysics, Astronomy and Mechanics, Department
of Physics, University of
Athens, Panepistimiopolis, GR-157 84, Athens, Greece
   \email{gpetrie@hao.ucar.edu,tsingan@phys.uoa.gr}\\
}

\date{received;accepted}

\abstract{

We analyzed a coronal loop observed with the {\it
Normal Incidence Spectrometer} (NIS), which is part of the {\it
Coronal Diagnostic Spectrometer} (CDS) on board the {\it Solar and
Heliospheric Observatory} (SOHO). The measured Doppler shifts and
proper motions along the selected loop strongly indicate
unidirectional flows. Analysing the Emission Measure Curves 
of the observed spectral lines, we estimated
that the temperature along the loop was about 380\,000~K.
We adapted a solution of the ideal MHD steady 
equations to our set of measurements. The derived 
energy balance along the loop, as well as the advantages/disadvantages
of this MHD model for understanding the characteristics of solar 
coronal loops are discussed.}

\maketitle


\section{Introduction}
Loops are one of the basic features of the solar corona. They are curved, 
filamentary features, connecting regions with opposite photosperic magnetic 
fields. They represent a basic component of active regions in the Extreme 
UltraViolet (EUV) and their understanding should contribute to the solution
of the coronal heating problem.

A specific class of loops are the so-called \lq cool EUV\rq\ loops,
with temperatures of $T < 10^6$~K (Bray \etal\ 1991). Their 
systematic study began with the analysis of {\it Skylab} data (Foukal, 1975,
1976), where it was stated that only the presence of flows could
explain the height reached in the corona by these cool structures.
The use of {\it Skylab} observations helped us understand that cool loops are 
different objects to the \lq hot\rq\ ones ($\geq 10^6$~K), since very often they 
are nearby but not co-spatial with them (Dere 1982, Habbal \etal\ 1985). 
Moreover, Sheeley (1980) noticed a large time variability of the cool loops 
in contrast to the hot ones.

After the launch of SOHO, a number of studies on active
region loops were based on observations from CDS/NIS and from the
{\it Extreme Ultraviolet Imaging Telescope} (EIT) in the EUV. 
In a study of five active regions, observed with CDS/NIS, Fludra \etal\ (1997) 
confirmed the frequent morphological distinctiveness of the cooler loops.
Moreover, observations in the \ov, 630~\AA\ line, again with CDS/NIS,
showed that cool loops present Doppler shifts all along their length of 
the order of $\pm (50-60)$~\kms\ that, in some cases, can reach values 
of $\simeq 200-300$~\kms\ (Brekke 1999, Kjeldseth-Moe \& Brekke 1998,
Brekke \etal\ 1997). Furthermore, their time variability can be as short as 3~minutes, 
as is deduced from movies obtained by the {\it Transition Region And Coronal 
Explorer} (TRACE), although some loops appear stable for up to about 3~hours.
It is believed that measurements of the temperature along loops will be able 
to discern their proper heating mechanism. However, temperature measurements 
carried out along $1.5-2.\times\, 10^6$~K
loops, using the ratio of the narrowband filters 171, 195~\AA\, from TRACE and EIT 
(Lenz \etal\ 1999, Aschwanden \etal\ 1999) seem in contradiction to the 
measurements carried out with spectral data from CDS/NIS (Schmelz \etal\ 2001). 
A lot of effort has been dedicated to the solution of this controversy (e.g. Martens \etal\ 2002,
Aschwanden, 2002) to clarify whether the low spatial resolution of CDS or the low 
spectral resolution of the narrowband instruments is to blame for a bias in 
the temperature measurements. Recently, Del Zanna \& Mason (2003) pointed out that 
part of the problem lies in the proper correction of the measured loop emission 
for the diffuse background emission along the line of sight (LOS).
They showed that the two emissions seem to originate from plasma in a different thermal state. 
Furthermore, they noticed that the response function of the 195~\AA\ TRACE narrowband 
was computed with out of date atomic data parameters, suggesting that the scientific results based
on the 171/195 ratio should be revisited.

However, the methods that are used to deduce plasma temperatures from narrowbands or spectral 
line intensities use assumptions that are sometimes difficult to justify and could 
lead to more than one solution, as is discussed by Judge \& McIntosh (2000).

Another approach could be to use a model for the computation of plasma parameters,
such as the temperature and to compare the computed intensity flux with the observations in
order to check whether the model can reproduce the observations (e.g. Aschwanden \etal\ 2000,
Reale \& Peres 2000, Del Zanna \& Mason 2003 and Petrie \etal\ 2003).

The first loops succesfully described by a model were the very hot 
($3\times\ 10^6\,-\,10^7$~K) X-ray ones for which the hydrostatic assumption 
holds (Rosner \etal\ 1978). 
However it seems that the less hot ones ($10^6\,-\, 2.\times\ 10^6$~K), observed 
by EIT and TRACE, are not in hydrostatic equilibrium in general (Aschwanden \etal\ 2001, 
Winebarger \etal\ 2003). The hydrostatic scale height at $10^6-2\times\ 10^6$~K is 
smaller than the height usually reached by these loops.

An attempt to model the intensities and flows along cool loops (Peres 1997), 
observed in the \ov\ 630~\AA\ line by CDS, concluded that steady hydrodynamic siphon 
flows could not explain the loop apex brightness. The same author, using a time-dependent 
calculation, suggests that these discrepancies could be a result of the impulsive nature 
of the heating. The disadvantages of a hydrodynamic unidirectional flow approach to 
explain TRACE loop observations are described in Patsourakos \etal\ (2004).

However, Petrie \etal\ (2003) showed that steady flows can be more successful
in reproducing loop measurements if one takes into account self consistently the 
magnetic forces in the momentum equation.
They were able to match the measured electron densities,
temperatures and flows along the loops with the results from a steady 2D MHD
model. Moreover, it was concluded that flows have an important
influence on the loop energy balance. This means that flows, along with
temperature and density, are physical quantities that should
be carefully determined in order to proceed in constraining the
loop heating function. This is in contrast to the common belief 
that plasma velocities do not affect the heating balance of a coronal 
loop significantly and therefore static models are adequate to 
determine the heating function, since coronal plasma kinetic
energies make up only a small proportion of the total energy budget.
\begin{figure*}[!ht]
\resizebox{0.9\hsize}{!}{\includegraphics*{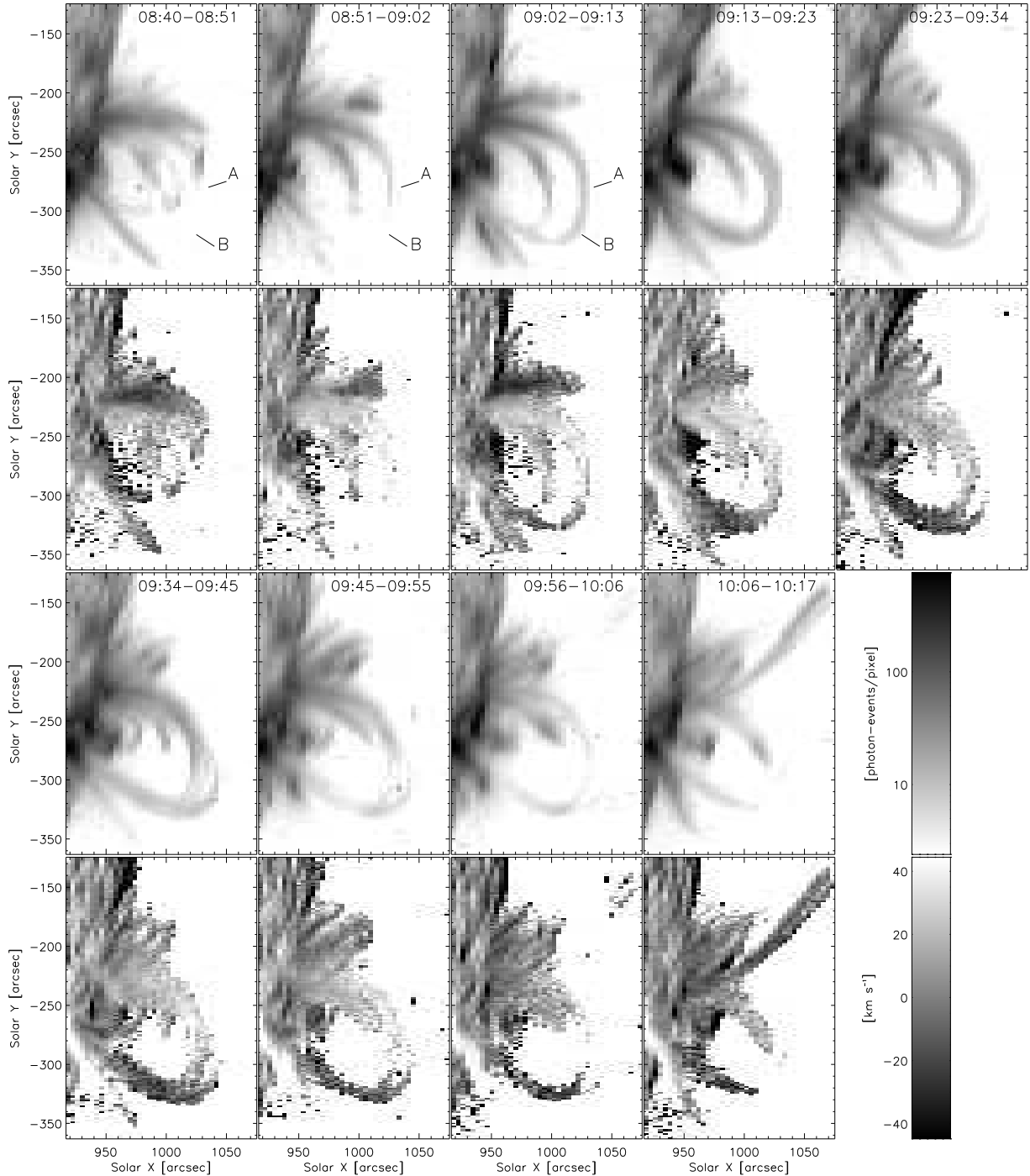}}
\caption{Total line intensity maps (first, third rows) and the
corresponding Dopplergrams (second and fourth rows) in \ov\ 630~\AA\ 
(maximum ion concentration at 250\,000~K), showing the evolution of the active region.
The segments A and B in panels 2 and 3 show the selected loop. The
loop evolution in panels 2 and 3 of the first row and the corresponding
Dopplergrams indicate that the loop is replenished from
its northern footpoint by a unidirectional flow. The first row
images indicate proper motions of the order of $30\pm 6$~\kms. The pattern of 
the loop remains roughly unchanged from 09:02 to 09:55~UT. The loop is fainter in 
the 09:56-10:06~UT raster and disappears in the 10:06-10:17~UT one. The Dopplergrams show 
a constant flow pattern along the loop for all its life time. We used a logarithmic 
scale for the intensities to enhance the loop contrast.}
\label{rasters}
\end{figure*}

In the present study we selected an active region loop with
time scales of about 1 hour and a dominant unidirectional
flow for comparison with the MHD model. Note that this is 
a cool loop, observed in transition 
region lines, with temperature much lower than the
$\simeq 10^6$~K studied in Petrie \etal\ (2003). 
The emphasis in this paper is on a detailed presentation of 
our measurements, since the MHD
model was already presented in Petrie \etal\ (2003). Thus, 
Sections 2 $\&$ 3 are devoted to the observations and their critical 
analysis, while in Sect. 4 our results are fitted to the MHD model. 
The advantages and disadvantages of our modelling technique are briefly and 
critically discussed in Sect. 5, while our main conclusions are given 
in  Sect.~6. 

\section{Observations, data reduction and measured flows in \ov\ 630~\AA}
\label{observations} 
The NIS spectrometer on CDS simultanously records two spectral bands:
the first (NIS 1) covers the 308-381~\AA\ part of the solar spectrum and the second 
one (NIS 2) the 513-633~\AA\ part. The spectral resolutions are respectively 0.32~\AA\ 
and 0.54~\AA. Usually, only some selected parts of the spectrum 
(e.g. some spectral lines plus their adjacent continuum) are kept to speed up the 
tranfer of the data to the ground.
Each single exposure records a field of view of a 240\arcsec\ long, 
North-South oriented slit. Square fields of $4\rq\ \times\ 4\rq$ are obtained by 
activating a scan mirror in the East-West direction.
\begin{figure}[!hb]
\begin{center}
\resizebox{1.\hsize}{!}{\includegraphics*{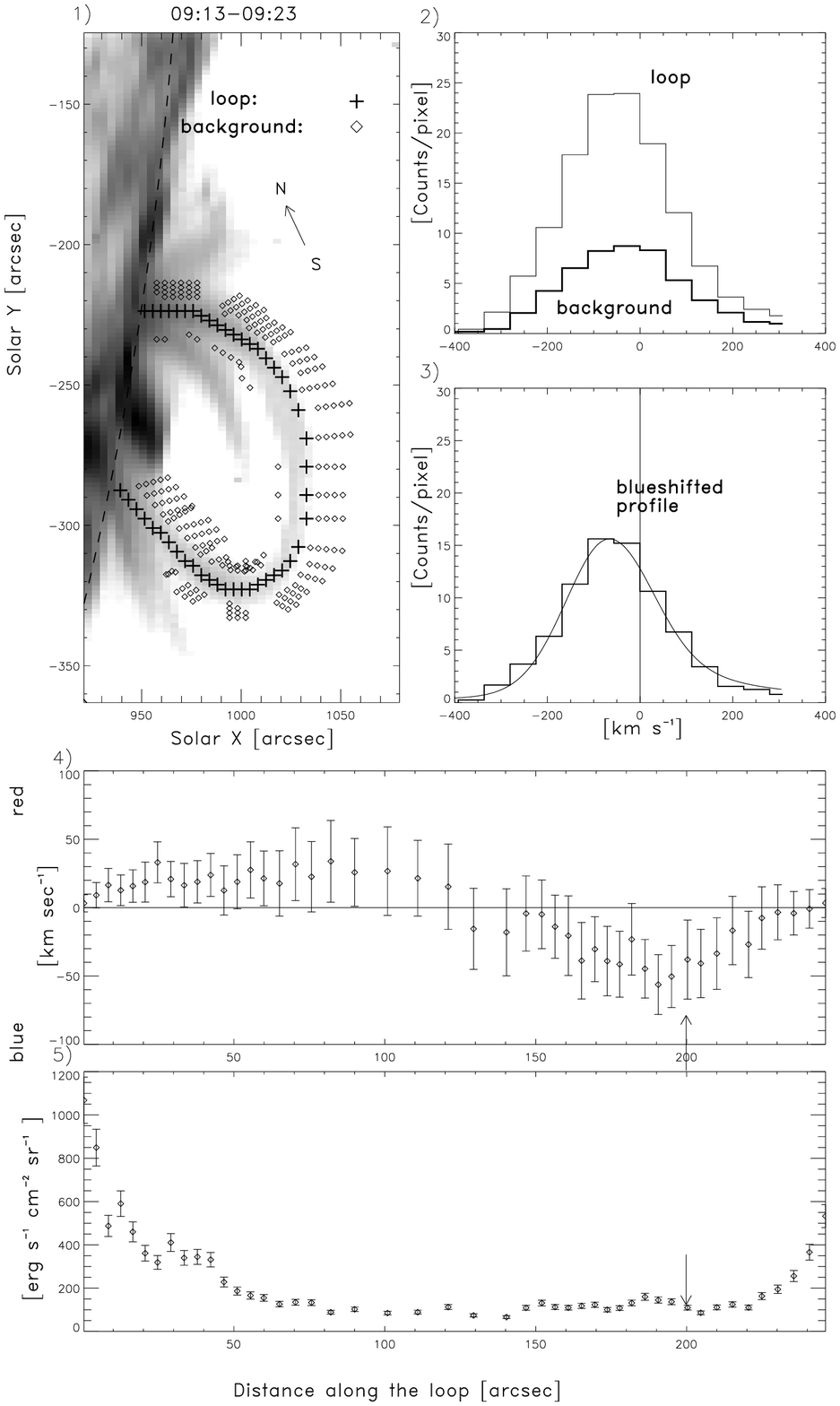}}
\end{center}
\caption{A thorough study of the loop in \ov\, 630~\AA. In
panel~1, crosses are the sampling points along the loop and
diamonds are the corresponding background ones. Panel~2 shows one
spectral profile taken on the loop and the corresponding
background one (dispersion axis is in \kms). Panel 3 is the loop
profile of panel 2 after the subtraction of the background profile. 
The smooth line is computed by the fitting procedure and estimates the 
total intensity and Doppler shift. Panel 4 shows 
the Doppler shift along the loop (starting from the northern footpoint). 
Panel 5 shows the flux emitted from the loop plasma along the loop in
physical units. The two arrows in panels 4, 5 indicate the Doppler shift 
and intensity values derived from the individual spectral profile of panel 3.
\label{back_vel_int}}
\end{figure}
On October 26 and 27, 1999, CDS/NIS observed
the active region NOAA~8737 on the S-W limb of the Sun, executing
218 scans with an 11~minute cadence (Fredvik \etal\ 2002). 
Each slit exposure simultaneously recorded six spectral lines emitted
by the ions \hei, \oiii, \ov, \nevi, \mgix, and \fexvi\ at the
wavelengths 584~\AA, 599~\AA, 630~\AA, 562~\AA, 368~\AA\ and
361~\AA, sampling the solar atmosphere in temperatures of the
upper chromosphere, the transition region and the corona. The
active region was also observed by SUMER (Fredvik \etal\ 2002) and
Yohkoh, which recorded a flare.

We selected the CDS data taken between 8:40 and 09:23~UT and for their
treatment, we corrected the CCD bias, flatfield, burn-in effects 
and cosmic rays (see Del Zanna 1999 for an introduction to 
CDS data reduction). From each individual spectral profile of the 
\ov\ 630~\AA\ line, we subtracted the adjacent pseudo-continuum, formed 
by the scattered light from the solar disk (see Del-Zanna \& Mason 2003).
We also corrected the mis-alignement between the two spectral bands
of NIS, using the {\sc nis\_rotate} routine.

Then, we applied a fitting routine to these profiles to
produce intensity and Doppler shift maps. As the
images were taken after the SOHO recovery, we had to take into
account the change of the PSF of the spectrometer, which produces
broadened line profiles that differ from a purely Gaussian 
shape (Thompson, 1999). The Doppler shift maps were corrected for 
the geometric distortion. As there is no wavelength calibration 
on-board, we made the assumption that the Doppler shift is equal 
to zero at a quiet region very close to the limb, because there
the Doppler shift of \ov\ 630~\AA\, has statistically very small
values (see in Peter \& Judge 1999 their Fig~4). Figure~\ref{rasters} 
presents nine consecutive intensiy images, presenting the 
loop from its birth to its disappearance. Each row of intensity 
images is followed by the row with the corresponding Dopplergrams.
In panels~1 and~2 of the first row, the loop segment AB 
is filled from the northern footpoint whereas in panel~3, it is full of 
bright plasma. Moreover, in the Dopplergrams, the north footpoint 
is redshifted contrary to the south one. These facts could be 
interpreted as material flow from the north footpoint to the south 
one and, because of a possible inclination of the loop plane away 
from the observer, we observe these values of Doppler shifts along its length.
The loop is visible from 09:02 to 10:06~UT and disappears 
in raster 10:06-10:17~UT. The corresponding Dopplergrams show that the velocity 
pattern along its length is the same during this time.

In an effort to study the loop physical parameters, we sampled the loop in the
09:13-09:23~UT raster (see Fig.~\ref{back_vel_int}). For each spectral profile
selected along the loop, we chose up to 4 more profiles nearby,
but outside the loop (Fig~\ref{back_vel_int}. panel 2). We subtracted the mean of
these nearby profiles from the loop ones, to correct for the
background effect. The background in our case is composed of
photons that are emitted along the LOS but
outside the loop (the plasma is optically thin). The fitting of
the corrected profiles provided the Doppler shift and the total
intensity along the loop. These measurements are presented in 
Figure~\ref{back_vel_int}, panels 4,~5, along with errorbars, derived by 
the fitting process. Even with the large error bars, a clear 
variation of the Doppler shift from red (positive velocities) to
blue shift is present.

In panels 4 and 5 of Fig~\ref{back_vel_int} the total projected length of the 
loop is 245\arcsec. To estimate the maximum length of the loop, we used the MDI magnetogram of 
24/10/1999, where the studied active region (NOAA~8737) has a maximum horizontal length of 
$\simeq\ 200$\arcsec. Supposing that the loop is semicircular and, in a rather extreme case, 
its footpoint separation is of the size of the active region, we estimated that its maximum 
length is about 314\arcsec\ which corresponds to 225~Mm. This estimation assumes that the 
loop connects two opposite polarities from the same active region. This must be the case 
since NOAA~8737 is isolated and far from other active regions visible on the magnetogram. 
The loop length will be used for the calculation of the Alfv\'en travel time in 
section~\ref{model}.

\section{Emission measure analysis along the loop}
\label{EML}
\begin{figure*}[!ht]
\resizebox{.85\hsize}{!}{\includegraphics*{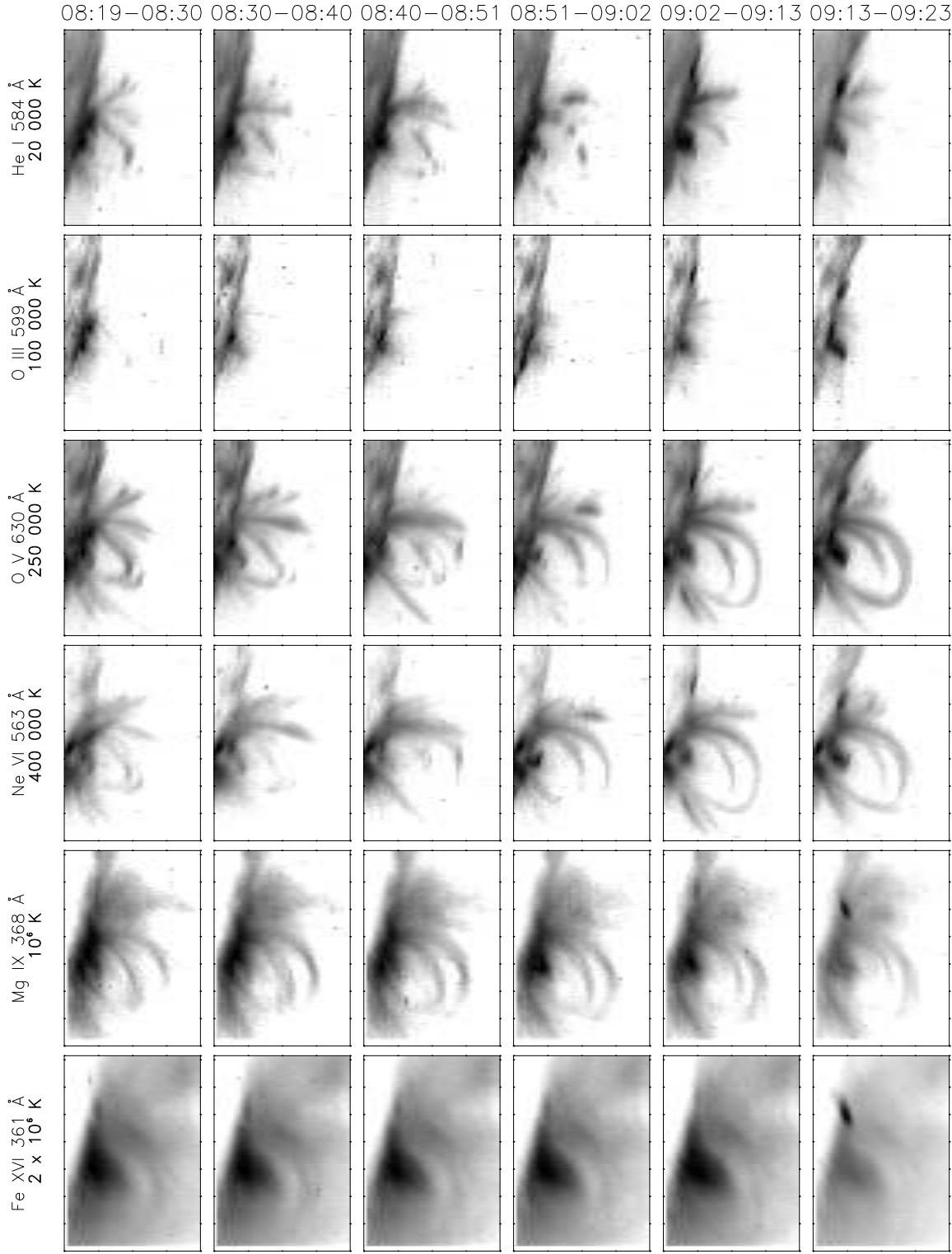}}
\caption{Images from all spectral lines, observed during 6 rasters, ordered by increasing
temperature of maximum ion concentration. The \mgix\ and \fexvi\ images co-alignment was corrected 
to account for the spatial offset between NIS-1 (308-379~\AA) and NIS-2 (513-633~\AA) parts of 
the CDS detector. The loop seems co-aligned in the \ov\ 630~\AA, \nevi\ 563~\AA\ and \mgix\ 
368~\AA\ spectral lines. However, the loop evolution is similar in the \ov\ and \nevi\ 
lines, while this is not the case for the \mgix\ line. As in Fig~\ref{rasters} the unidirectional 
flow is evident by comparing the individual images in \ov\ and \nevi.}
\label{all_ions}
\end{figure*}
\begin{figure}
\resizebox{1.\hsize}{!}{\includegraphics*{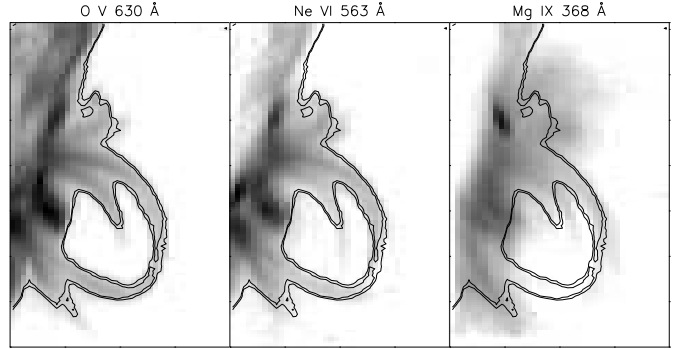}}
\caption{ Intensity images of the \ov\ 630~\AA, \nevi\ 563~\AA\ and \mgix\ 368~\AA\ lines, 
(09:13-09:23~UT) with the contours of \ov\ 630~\AA\ line superimposed. Whereas the \ov\
contours match well the  \nevi\ image, they match only part of the \mgix\ loop.
Furthermore, the \mgix\ loop apex seems shifted to the right relative to the \ov\ contours. }
\label{contours}
\end{figure}

\subsection{Estimation of the loop temperature}
\label{uncertain}
In Figure~\ref{all_ions}, the images of the six spectral lines recorded during 6
rasters, from 08:19 to 09:23~UT, are presented. We wanted to select the spectral lines
in which the loop has the same morphology and time evolution as in \ov\ 630~\AA.
We noticed that the loop morphology in the \ov\ 630~\AA\ and \nevi\ 563~\AA\ lines
is almost identical in the 6~rasters. In the 09:13-09:23~UT raster that we analysed, the \mgix\ 
line image has a loop which is partly cospatial with the \ov\ and \nevi\ one but with 
a less similar morphology (Fig.~\ref{contours}). Furthermore, the time evolution of the \mgix\ loop 
is different from \ov, \nevi\ ones, as it can be seen in the 6~rasters. On the other hand, we cannot 
see the loop in \oiii\ and in \hei\, only its footpoints could be guessed. In the hotter 
\fexvi\ 361~\AA\ line, no loop features seem co-spatial with the \ov\ ones.

Therefore, it seems that the considered loop plasma emits mostly in
the temperature range of \ov\ 630~\AA, \nevi\ 562~\AA\ 
lines, if we take into account the loop morphology and time evolution. 

We will include the \mgix\ 368~\AA\ spectral line further in the analysis, but not
in the estimation of the loop temperature for the reasons mentioned above as well as for those 
discussed in the following paragraphs.

Similarly to the case of \ov, we computed the total intensity for the \nevi, \mgix\ lines along the loop. 
We derived the contribution functions $G(T)$ for the three spectral lines, with the CHIANTI package (Dere
\etal\ 1997) using a constant electron density of $N_e\,=\, 10^9 cm^{-3}$,
considering a hybrid abundance for the solar corona (Fludra \& Schmelz 1999). Assuming that the plasma is in
ionization equilibrium, we applied the Mazzotta \etal\ (1998) ionization fractions. For each 
point along the loop and for the three spectral lines (\ov, 630~\AA, \nevi\ 562~\AA, \mgix\ 368~\AA), we
computed the ratios $I/G(T)$, where $(I)$ is the total line intensity. 

In Fig~\ref{loci}, panel~1, each of these ratios is presented, for a single point along the loop,
as double dashed curves, in order to account for uncertainties that will be discussed in Sect.~3.2.
An emission measure distribution of the loop plasma along the LOS, sharply peaked at $380\,000$~K, 
at the crossing point of the \ov\ and \nevi\ curves, can reproduce the loop brightness in these 
two spectral lines. A similar sharp distribution that would reproduce the loop brightness
in the three lines (\ov\ 630, \nevi\ 562, \mgix\ 368~\AA) should peak at $\simeq\ 500\,000$~K, 
but would need an electron density of $10^{12}$~cm$^{-3}$. Such a high density cannot 
be supported in such a gradually stratified manner as implied by the flux profiles over the
timescales observed.

Therefore, we considered that the plasma temperature of the loop is defined at the crossing 
point of the \ov\ and \nevi\ curves (Fig~\ref{loci}). The temperature estimated with 
this method is almost constant along the loop, of the order of $380\,000$~K. 
We did not include the \mgix\ measurement in the loop temperature estimation because we think 
that it originates from a different loop. This choice is supported by the fact that there 
is a general belief (Fludra \etal\ 1997) that cool loops (i.d. \ov, \nevi\ ones) are different
to the hot ones (seen in \mgix\ lines in this case).

\subsection{Uncertainties in the temperature estimation}
The computations of the G(T) functions include some assumptions that need to be 
discussed. For example, ion populations at the upflowing loop footpoint, where plasma 
probably flows across strong temperature gradients, should shift from the assumed 
ionization equilibrium values (see Spadaro \etal\ 1991). The time needed for the 
underabundant \ov, \nevi\ ions to reach their equilibrium values can be estimated as 
the inverse of the ionization rates of \oiv, \nev\ respectively. We estimated these 
quantities from the corresponding table of Shull \& Van Steenberg (1982), considering 
a temperature of $380\,000$~K and an electron density of 10$^9$~cm$^{-3}$. The 
lifetimes we found are 2~sec and 17~sec respectively. If we suppose a flow speed of 
about 50~\kms, at the upflowing footpoint, the equilibrium values of the populations 
will be reached within the first 1.5\arcsec\ from the upflowing footpoint of the loop. 
This value is a very small fraction of the loop length and therefore we can 
neglect this effect in our estimations. 
\begin{figure}[!t]
\begin{center}
\resizebox{.8\hsize}{!}{\includegraphics*{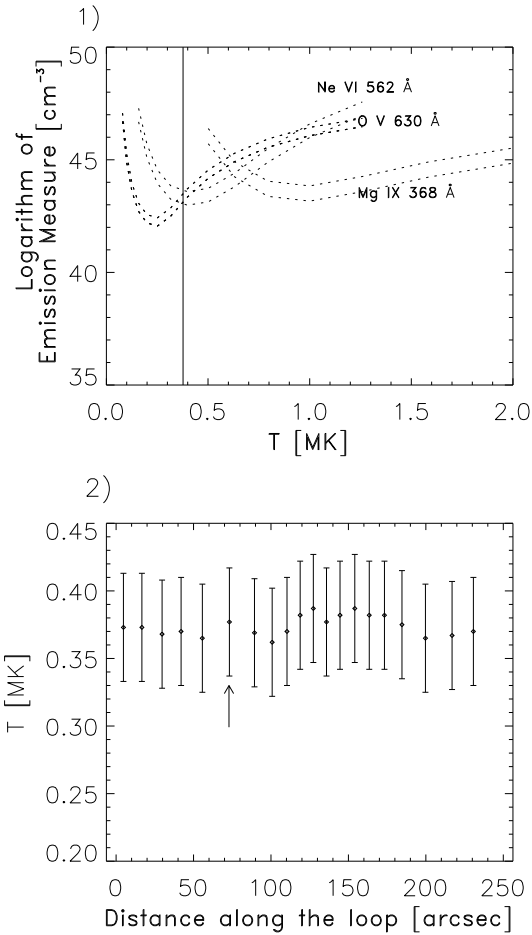}}
\end{center}
\caption{In panel~1, the
emission measures for \ov\, 630~\AA, \nevi\ 562~\AA\ and \mgix\ 368~\AA\ for a given part of the loop are 
plotted as functions of temperature. The loop plasma temperature is estimated at the crossing point between
the \ov\ and \nevi\ curves. The dashed lines show the effect of the uncertainties discussed in
Section~\ref{uncertain}. They produce two extreme crossing points for the emission measure curves and are
translated to error bars on the temperature. Panel~2 shows the derived temperature along the loop. 
The arrow points to the temperature measurement presented in panel~1.}
\label{loci}
\end{figure}
\begin{figure} 
\begin{center}
\resizebox{.9\hsize}{!}{\includegraphics*{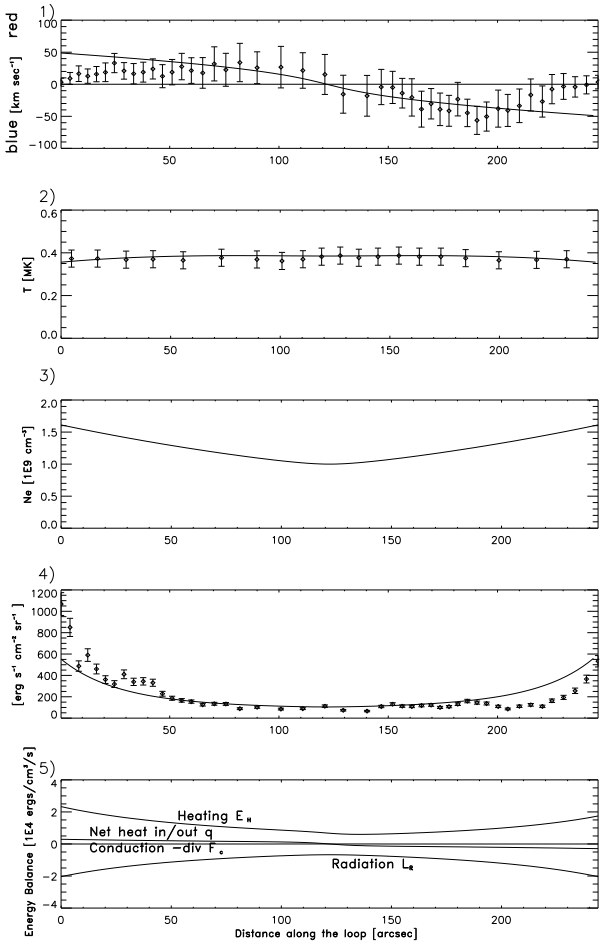}}
\end{center}
\caption{A summary of the measurements we carried out along the loop,
fitted with the MHD model. The data are represented as diamonds with 
their error bars and the MHD computations are the full lines. Here, 
the modelled flow is projected along the LOS so that we can compare 
with the data. In panel 2, the temperatures were constrained using the 
Emission Measure curves (see Fig.~\ref{loci}). Panel~3 shows the 
computed electron density along the loop. The computed temperatures 
and electron densities are used to compute the intensity in the \ov\ 
spectral line, which is compared with the observed one in panel~4.
In panel~5, where we do not
have any data measurements, we present the terms of the energy
balance along the loop. The heating $E_H$ is the resulting energy
from the balance of the other terms. Note that the conductive 
losses are very small in the scale of panel~5.}
\label{summary}
\end{figure}
Another case of departure from ionization equilibrium can hold if the loop is subject to time variations
of the temperature along its length, due to heating or cooling events (see for example the simulations performed 
by Bradshaw \& Mason, 2003a, 2003b). In the case of our loop, it seems that such variations could have
occurred during the \lq birth\rq\ of the loop, 20~minutes before the period that we analysed 
(see Fig~\ref{rasters}). Following the previous reasoning, we found that the longest timescale,
computed from the ionization and recombination rates for the two cooler ions, is the recombination 
rate of \ov, which is roughly equal to 1~minute. From the previous simple calculation we think 
that the ionization equilibrium holds for \ov\ and \nevi\ in the measurement we carried out.

The fact that the available spectral lines originate from different elements 
brings uncertainties that are due to the element abundances. Schmelz \etal\ (1996) studied 
a sample of 33 non-flaring active regions and concluded that the Ne/O ratio present a variation of 
a factor~4. Moreover, the CDS calibrated intensities have an accuracy of the order of 30\% 
(Del Zanna \etal\ 2001). Taking into account these facts, we estimated that the error bars 
on the temperature are of the order of $\pm 40\,000$~K (see Fig.~\ref{loci}).

\section{Comparison with the MHD model}
\label{model}
The observed loop is modelled using steady 2D MHD equilibrium
solutions with compressible unidirectional flow (Petrie \etal\ ~2002,
2003). The details of the modelling technique, including
a self-consistent heating model, are described fully in Petrie \etal\
(2003). 
The use of a steady model to describe the observed loop
is justified since the time scale involved in the observed
loop (roughly 1~hour as seen in Fig.~\ref{rasters}) is significantly larger 
than the time for a disturbance to travel along the loop with 
an Alfv\'en speed of 1000 \kms\ (less than 4~minutes as the loop
length is 225~Mm, see section~\ref{observations}).

Moreover, for a radiative loss rate at $380\,000$~K of 
$\Lambda\ (T) \approx 10^{-10.4} T^{-2}$ (Rosner \etal\ 1978) and 
an electron density of $10^9$~cm$^{-3}$, the cooling time is
$\tau\,=\, \frac{3\,k_B\,T}{n_e\,\Lambda(T)}\approx 10$~minutes. 
This is considerably smaller than the loop lifetime. Therefore we 
consider that the loop is in thermal equilibrium.

Given the loop dimensions, the model computes the flow, the mass
density, temperature and heating along the loop, for a calculated shape
of the magnetic field. 
The computed loop geometry is orientated in space by tuning two angles, 
so that the loop projection on the plane of the sky fits the observed loop shape 
(see Petrie \etal\ 2003 for further details). 
One more constraint to this operation is that the computed line of sight 
velocity should fit the observed Doppler shifts (see Fig.~\ref{summary} panel~1).

From the computed electron densities and temperatures, we calculated an \ov\ 
intensity flux along the loop, which was compared to the observed flux 
(see Fig.~\ref{summary} panels~2,3,4). This forward process supposed that: 

\begin{itemize}
\item The loop differential emission measure along the LOS is 
described by a Dirac function centered at the computed temperature
\item The loop width is constant and equal to 4~Mm along the loop
\item The filling factor is taken to be equal to 1 but our model is 
qualitatively unchanged for filling factors down to the order of 0.5.
\end{itemize}

From Fig.~\ref{summary} we see that the model succesfully represents the 
smooth drop of the intensity along the loop. Recall that hydrodynamic 
siphon flow calculations do not produce the large intensity scale heights 
along \ov\ loops observed at the limb by CDS except if an ad hoc constant temperature 
is imposed along the modeled loop (Peres 1997).

In panel~5, we present graphs of the functions 
involved in the energy equation. The radiative term ($L_R$), being 
negative to represent an energy loss, dominates the energy balance, 
and conduction (div $F_c$) is negligible because the model is 
nearly isothermal. The net heat in/out (q) is computed from the 
first law of thermodynamics, (see Petrie \etal\ 2003 Eq. 3). 
Due to its dependence on the flow it is anti-symmetric. From the 
summation of the above mentioned energy terms we compute the 
unknown heating distribution along the 
loop ($E_H$) which safeguards energy conservation. From Fig~\ref{summary}
we see that $E_H$ has a minimum near the loop top, local maxima at the 
footpoints and is stronger at the upflow footpoint. 
Moreover, we deduce the heating function
without any assumption about which mechanism is producing this
heating, simply from the energy conservation.

\section{Discussion}
\label{discussion} Coronal loops are almost always modelled using
hydrostatic or hydrodynamic solutions, i.e. solutions that do not
explicitly include the magnetic field. However, in real plasma
systems the magnetic field inevitably causes the plasma to depart
from one-dimensional equilibrium, particularly in such a
magnetized, sparse medium as the solar corona. Even
near-isothermal images of near-static active regions show
disagreement with hydrostatic equilibrium in their characteristic
density scale height and complex structure (Aschwanden \etal\ 2001, Fig.~8).
The graph of density against arclength of a field line in a 
hydrostatic plasma is likely to differ from that of a field line 
of identical size and shape in a non-force-free magnetohydrostatic
plasma of the same characteristic temperature. This is reflected 
in the fact that the isobars and isotherms of non-force-free MHD 
models are generally very different from the isobars and isotherms of
hydrostatic equilibrium, which are parallel to the photosphere. The
physical parameters along a steady MHD loop can differ from hydrostatic
equilibrium because the cross-field compressive forces are generally not
distributed along the loop in the same manner as the usual field-aligned
hydrostatic forces (Petrie et al.,~2003).  The
justification for excluding this direct participation of the
magnetic field in traditional hydrostatic and hydrodynamic models
is that the Lorentz force is perpendicular to the magnetic field
and therefore does not influence the statics/dynamics along the
loop. However in 2D and 3D magnetic structures the Lorentz force
communicates across the loop with the pressure gradient,
gravitational and inertial forces, giving rise to the various
scale heights found in near-isothermal images.

In the study by Aschwanden et al.~(1999), EUV loops observed by
EIT on the disc are fitted with hydrostatic models. With one small
exception, only parts of loops were amenable to measurement and
modelling. Hydrostatic models were successfully fitted to these
loop parts. Then, Aschwanden et al.~(2001) selected entire TRACE
loops on the limb and concluded from hydrostatic modelling that
the observed scale heights were super-hydrostatic. 

Furthermore, the cases of flat temperature and density profiles measured 
along loops, observed by imagers as well as by CDS, cannot be reproduced
by steady hydrodynamic flow models (Patsourakos \etal\ 2004, Landi \& Landini 2004). 
Such tests should be repeated for loops where the measured temperature decreases 
toward the footpoints (e.g. Del Zanna \& Mason 2003).

However, MHD physics produces steady 
loop solutions that can easily fit the temperature and density 
profiles deduced by $\simeq 10^6$~K loops (Petrie \etal\ 2003).
Therefore, it appears that MHD steady flows may be useful
for the study and understanding of coronal loops, despite their 
limitations which we discuss in the following. 

One limitation of our modelling technique, compared with
other techniques, is that the loop cross section increases
from the footpoint upward, contrary to the observational
evidence. This is a consequence of the self-similarity
technique invoked in order to generate the analytical
solutions (Petrie \etal\ 2002): any two field lines can only
differ by a vertical translation. It is not obvious how to
re-derive the solutions maintaining some freedom in the loop width
function. Various 1D techniques allow free width functions by
ignoring the cross-field physics. In two dimensions we
do not have this luxury. Instead, we can interpret our results
taking this limitation into account.

Because the model loop legs are nearly vertical, our modelled loop
is much narrower at the footpoints than at the apex. 
Thus, the model overestimates the plasma velocity near the
footpoints, (c.f., top panel in Fig.~\ref{summary}). A model with
a nearly constant cross-section and the same mass flux would
better reproduce the decrease in velocity magnitude seen in
the observations.

As a consequence of the smaller width of the loop at the 
footpoints, the model over-estimates
the magnetic field strength, which in turn affects the plasma
parameters, e.g. in the present model the gas pressure and the
temperature drop. While the model is reasonable over most of the
loop length, it is weaker in the sections within 40\arcsec\ of
each footpoint. This is a small price to pay for a reasonable full MHD 
description over most of the loop length.

A trace of the enhanced heating at the upflow leg can be seen in
the enhanced line width that the \ov\ line presents at the loop
legs. However, the instrument broadening of the line profiles is
very large for CDS, so we cannot use this as a hint. Another
indication could be the fact that the north leg, is near a bright
region, visible in all images (see Fig.~\ref{all_ions}) which had a strong
brightening during the flare observed by Yohkoh.

TRACE observations, due to the higher spatial resolution, revealed 
the multistrand structure of loops, which has not been taken into account
in the present study.
Various attempts have been made to model loops as bundles of many
independent strands, each of them in hydrostatic equilibrium, by
simply superposing solutions (Aschwanden et al.,~2000; Reale \&
Peres,~2000). A major problem with this multistrand approach, however, is
that, while the hydrostatic equation is linear in its physical
parameters, the energy equation is highly nonlinear (as are the
hydrodynamical and MHD equations) so that two independent
solutions do not add together to form a new solution, in general.
However, Sakai \& Furusawa~(2002) did take into account this nonlinearity in a 
multi-thread model with flow, using a full 3D MHD code, but they do not include 
the loop curvature in their model.

\section{Conclusions}
We have compared detailed observations of flows and intensities in a cool solar 
loop with a 2D-MHD model.
We presented evidence that the plasma material is feeding the selected loop 
from the north footpoint, as one can see from the first~three intensity images in Fig.~\ref{rasters}. 
This, in combination with the Doppler shifts (Fig.~\ref{back_vel_int}), 
strongly indicates a unidirectional flow along the loop. The temperature 
has been also estimated, at first order, to be around $380\,000$~K
and seems to remain nearly constant along the loop length.

The three data sets (intensity, velocity and temperature),
along with the loop geometry, represent a sufficient constraint
for the free parameters of the MHD model. From the resulting
fitting process, we computed the energy terms that participate in
the energy balance and we deduced the shape of the heating
function along the loop. The heating function is stronger at the
footpoints. Moreover, the heating function presents an asymmetry,
being stronger at the upflowing leg. This may be a serious constraint
for modelling the mechanism that produces this heating function;
nevertheless, this is beyond the scope of this work. 
The measure of the non-thermal broadening of the spectral line along 
the loop used to deduce the flows (\ov\ in our case) can give 
information on the shape of the heating function 
(see also Spadaro \etal\ 2000). In our case however, due to the 
degradation of the CDS instrument after the loss of SOHO, this 
was not possible.

The shape of the heating function is likely to generalise across 
all models constructed in this way. A model with symmetric plasma
parameter profiles and unidirectional flow will always have symmetric
radiative loss and heat conduction profiles, as well as an anti-symmetric 
net heat in/out function (first law of thermodynamics). For near-isothermal
loops, heat conduction does not contribute much to the energy equation and
the radiative losses are concentrated at the footpoints in any model with
gravitational stratification. The heating balance will therefore balance a
combination of a symmetric radiative loss function and an antisymmetric
net heat in/out function. Such a combination can only give an asymmetric
heating function, concentrated at the footpoints and biased towards the
upflow footpoint. This bias is determined by the radiative loss function
and the net heat in/out function.

Using a 3D MHD code, Gudiksen \& Nordlund~(2002) give a 3D numerical model
where photospheric motion followed by field-line relaxation results in
footpoint heating by magnetic dissipation that causes heated plasma to
fill the loop from one footpoint to the other. It is reassuring that
their approach also predicts that biased footpoint heating is consistent
with unidirectional plasma flow.

In our previous work (Petrie \etal\ 2003), we deduced the same 
characteristics of the heating function for three other 
loops. However, we had studied loops that 
were in the temperature range of ($1.-1.5\times\ 10^6$)~K, 
whereas in this case, the loop is much cooler.
More loops, with clear unidirectional flows, should be studied in order 
to fully explore this new mathematical tool.
\vspace{0.5cm}

\begin{acknowledgements}
CG and GP acknowledge funding by the EU Research Training Network PLATON, contract number HPRN-CT-2000-00153.
CG would also like to thank the Research Comittee of the Academy of Athens, the Costopoulos Foundation as well as
Drs S. Patsourakos and G. Del Zanna for their userful suggestions. We also thank T. Fredvik for reading a 
previous version of the manuscript, as well as the anonymous referee for the valuable comments
which contributed to a significant improvement of our paper. CHIANTI is a collaborative project involving
the NRL (USA), RAL (UK) and the 
Universities of Florence (Italy) and Cambridge (UK), while SOHO is a joint project of ESA and NASA.
\end{acknowledgements}

\end{document}